\documentclass[twocolumn, twocolappendix]{aastex631}
\usepackage{comment}

\usepackage{amsmath,amssymb}
\usepackage{hyperref}

\shortauthors{M. De Furio et al.}
\shorttitle{CWISEP J193518.59-154620.3: Y-Y dwarf binary}

\begin{document}


\title{Discovery of the Second Y+Y Dwarf Binary System: CWISEP J193518.59-154620.3}


\correspondingauthor{Matthew De Furio}
\email{defurio@utexas.edu}

\author[0000-0003-1863-4960]{Matthew De Furio}
\affiliation{Department of Astronomy, The University of Texas at Austin, 2515 Speedway, Stop C1400, Austin, TX 78712, USA}
\affiliation{NSF Astronomy and Astrophysics Postdoctoral Fellow}

\author[0000-0001-6251-0573]{Jacqueline Kelly Faherty}
\affiliation{Astrophysics Department, American Museum of Natural History, 79th Street at Central Park West, New York, NY 10024, USA}

\author[0000-0001-8170-7072]{Daniella C. Bardalez Gagliuffi}
\affiliation{Department of Physics \& Astronomy, Amherst College, 25 East Drive, Amherst, MA 01003, USA}

\author[0000-0002-2592-9612]{Jonathan Gagn\'e}
\affiliation{Plan\'etarium de Montr\'eal, Espace pour la Vie, 4801 av. Pierre-de Coubertin, Montr\'eal, Qu\'ebec, Canada}
\affiliation{Trottier Institute for Research on Exoplanets, Universit\'e de Montr\'eal, D\'epartement de Physique, C.P.~6128 Succ. Centre-ville, Montr\'eal, QC H3C~3J7, Canada}

\author[0000-0003-4636-6676]{Eileen C. Gonzales}
\affil{Department of Physics and Astronomy, San Francisco State University, 1600 Holloway Ave., San Francisco, CA 94132, USA}

\author[0000-0003-2102-3159]{Rocio Kiman}
\affiliation{California Institute of Technology, 1200 E. California Boulevard, Pasadena, CA 91125, USA}

\author[0000-0002-2387-5489]{Marc Kuchner}
\affiliation{Exoplanets and Stellar Astrophysics Laboratory, NASA Goddard Space Flight Center, Greenbelt, MD, USA}

\author[0000-0001-7519-1700]{Federico Marocco}
\affiliation{IPAC, Mail Code 100-22, California Institute of Technology, 1200 E. California Boulevard, Pasadena, CA 91125, USA}

\author[0000-0003-0548-0093]{Sherelyn Alejandro Merchan}
\affiliation{Department of Astrophysics, American Museum of Natural History, New York, NY, USA}
\affiliation{Department of Physics, Graduate Center, City University of New York, New York, NY, USA}

\author[0000-0003-4225-6314]{Melanie Rowland}
\affiliation{Department of Astronomy, The University of Texas at Austin, 2515 Speedway, Stop C1400, Austin, TX 78712, USA}

\author[0000-0002-6294-5937]{Adam C. Schneider}
\affiliation{United States Naval Observatory, Flagstaff Station, 10391 West Naval Observatory Rd., Flagstaff, AZ 86005, USA}

\author[0000-0002-2011-4924]{Genaro Su\'arez}
\affiliation{Department of Astrophysics, American Museum of Natural History, Central Park West at 79th Street, NY 10024, USA}

\author[0000-0003-0489-1528]{Johanna M. Vos}
\email{johanna.vos@tcd.ie}
\affiliation{School of Physics, Trinity College Dublin, The University of Dublin, Dublin 2, Ireland}
\affiliation{Astrophysics Department, American Museum of Natural History, 79th Street at Central Park West, New York, NY 10024, USA}

\begin{abstract}
We present the discovery of a companion to the Y-dwarf, CWISEP J193518.59-154620.3, the second Y-Y dwarf binary detected to date. Y-dwarfs are the coldest known free-floating objects ($<$ 500 K) and on average represent the lowest mass objects directly formed through turbulent fragmentation of a molecular cloud. Studying their multiplicity allows us to place strong constraints on the ability to form multiple systems of planetary masses and approaching the opacity limit of fragmentation. Due to their physical properties, Y-dwarfs also serve as analogs to gas giant planets. CWISEP J193518.59-154620.3 has been shown to have a unique methane emission feature in its near infrared spectrum at 3.326 $\mu$m, potentially indicative of auroral processes without a clear origin. CWISEP J193518.59-154620.3 was observed with JWST's MIRI in the F1000W, F1280W, and F1800W filters. We applied a point-spread function (PSF) fitting algorithm using empirically derived PSF models and resolve a companion in the F1000W and F1280W filters separated by 172 milli-arcseconds, 2.48 au assuming the distance of 14.43 pc. Using the ATMO2020 evolutionary models, we estimate a mass of 12-39 $M_{\rm Jup}$ for the primary and 7-24 $M_{\rm Jup}$ for the companion assuming an age of 1-10 Gyr for a mass ratio of 0.55-0.62, resulting in an estimated period of 16-28 years. It is unknown which component of this binary exhibits the methane emission feature. We also resolve known companions WISE J014656.66+423410.0B and WISE J171104.60+350036.8B using MIRI data and present their F1000W and F1280W photometry.

\end{abstract}


\section{Introduction} \label{sec:intro}

Y-dwarfs are some of the coldest objects that have been directly observed outside of the solar system and have effective temperature ($T_{\rm eff}$) $\lesssim$ 500 K \citep{Cushing2011ApJ...743...50C, Kirkpatrick2012}, although a specific classification sequence is still yet to be determined \citep{Beiler2024ApJ...973..107B}. There have been roughly 50 Y-type brown dwarf candidates that have been discovered all at close distances $\lesssim$ 50 pc, either serendipitously or in mid-infrared surveys like that from the Wide-field Infrared Survey Explorer (WISE) \citep{Kirkpatrick2024ApJS..271...55K}. Assuming similar age distributions for field brown dwarfs, Y-dwarfs on average represent the lowest mass objects that form directly through the star and brown dwarf formation process, i.e. turbulent fragmentation within a molecular cloud \citep{Boyd2005A&A...430.1059B}.

Due to their low temperature and isolation, they are useful testbeds for understanding giant planet atmospheres. These objects do not have hydrogen fusion occurring in their cores, and therefore they cool over time. This results in a mass and age degeneracy where old, high mass brown dwarfs have similar atmospheric features as young, low mass brown dwarfs of the same $T_{\rm eff}$ \citep{Burrows+2001, Saumon2008ApJ...689.1327S}. Therefore, it is crucial to obtain mass measurements and spectra of these types of objects to calibrate evolutionary models of brown dwarf atmospheres \citep[e.g.][]{Morley2012, Morley2014a,Morley2014b,Phillips+2020}. The only way to directly measure masses of these particular objects is through their presence in a multiple system and tracing their orbit \citep{Liu2008ApJ...689..436L, DupuyLiu2017}.

\begin{figure*}[h]
\includegraphics[width=\textwidth]{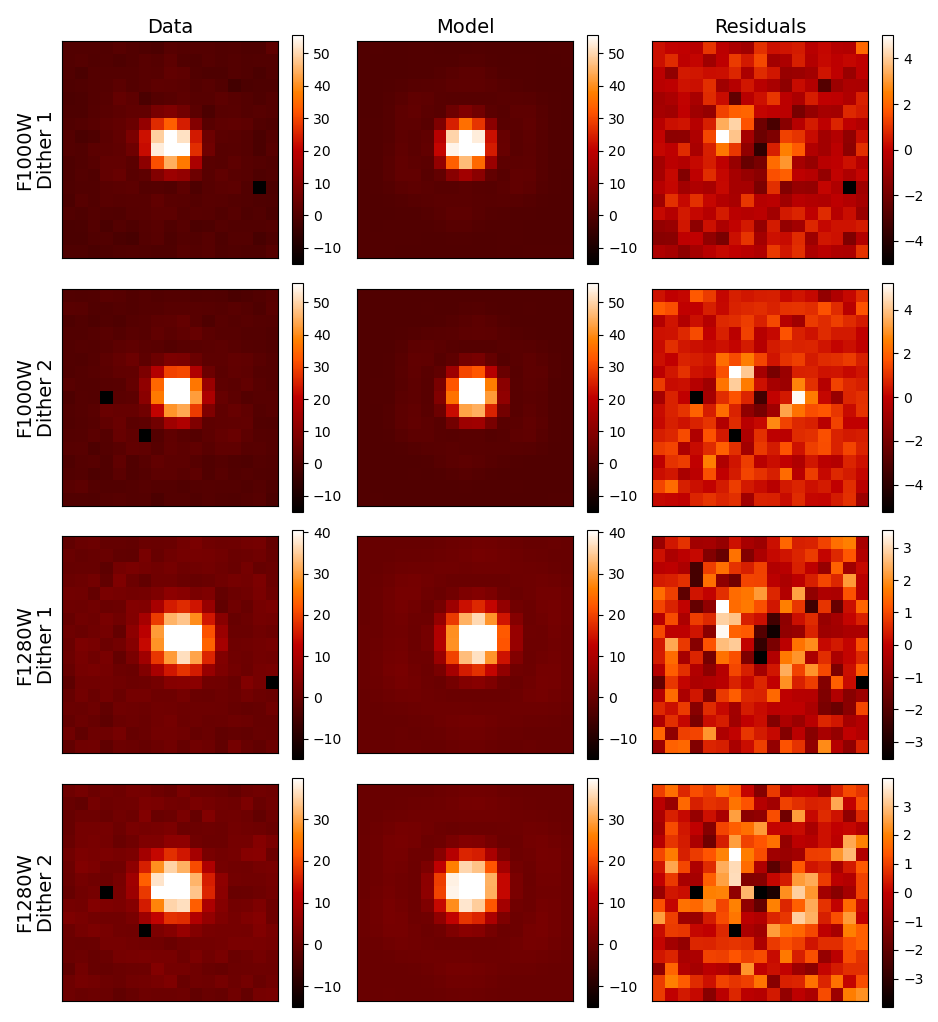}
\caption{Simultaneous single PSF fitting to F1000W and F1280W data of W1935. Left column shows a cutout of the data, middle column shows the PSF model, and right shows the residual. First row shows single-PSF fit for first dither in F1000W, second row for second dither in F1000W, third row for first dither in F1280W, and fourth row for second dither in F1280W. Single-PSF fit gives a $\chi_{\nu}^{2}$ = 2.3. Residuals follow expected pattern in core for a single-PSF fit to the photocenter of a binary, i.e. negative residuals in core (model overestimate of flux) and positive residuals immediately away from core (model underestimate of flux). Units are in counts/s. Each image has a size of 17 by 17 pixels.}
\label{fig:single}
\end{figure*}

\begin{figure*}[h]
\includegraphics[width=\textwidth]{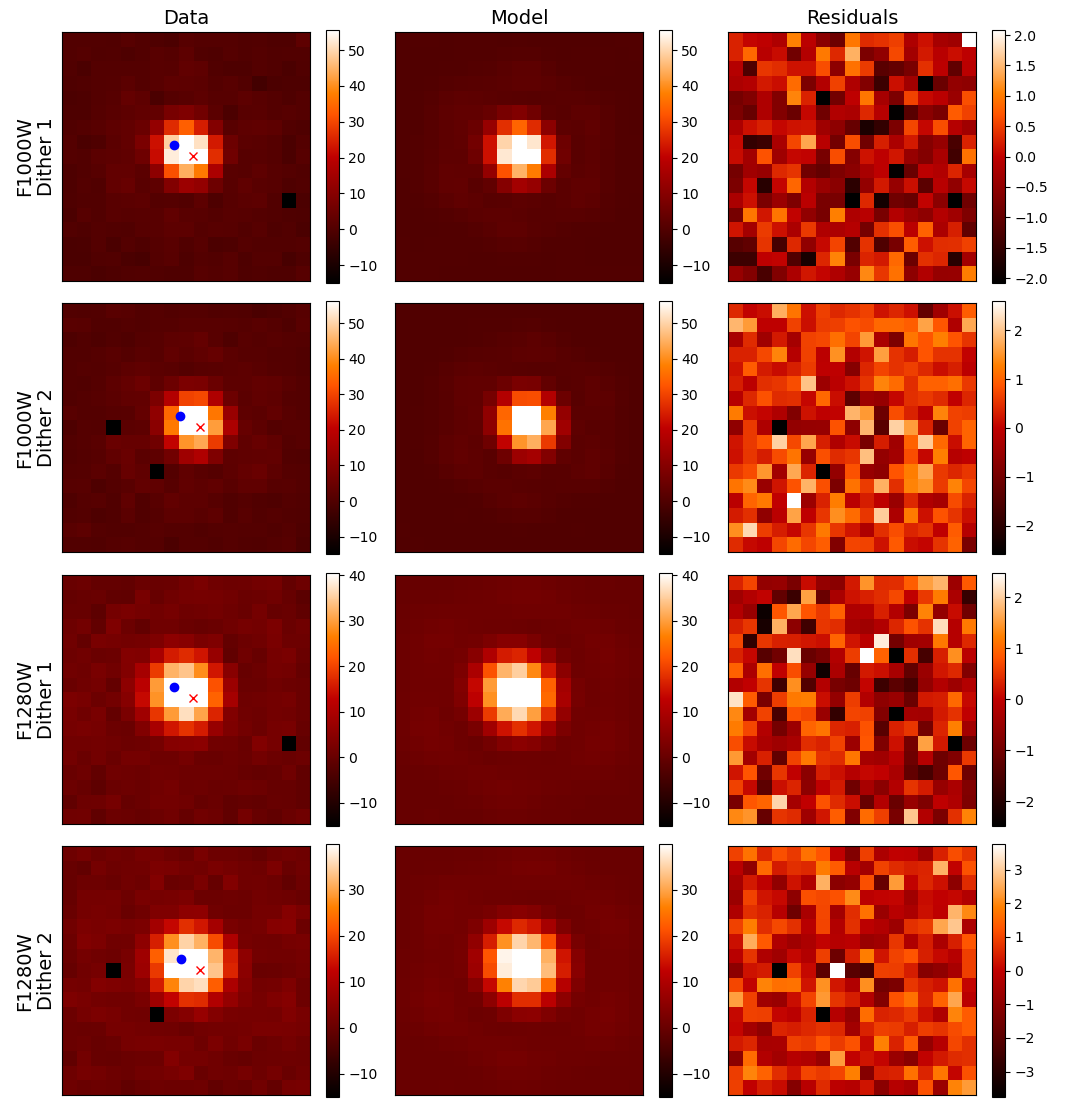}
\caption{Simultaneous double PSF fitting to F1000W and F1280W data of W1935, arranged as in Fig. \ref{fig:single}. Double-PSF fit gives a $\chi_{\nu}^{2}$ = 1.4. Residuals follow expected pattern for a good model fit to the data, i.e. random noise. The red x marks the location of W1935A, and the blue circle marks the location of W1935B.}
\label{fig:double}
\end{figure*}

Wide, planetary-mass companions to stars may have masses lower than those of known Y-dwarfs, e.g. Eps Ind Ab \citep{Matthews2024Natur.633..789M} and 14 Her c \citep{BardalezGagliuffi2025arXiv250609201B}, but their masses are difficult to directly measure due to their long periods. Close, planetary-mass companions to stars have shorter periods but are difficult to directly detect and obtain a spectrum due to the large flux contrast relative to the primary star, e.g. HD 206893 c \citep{Hinkley2023A&A...671L...5H}. On the other hand, close brown dwarf binaries are much easier to directly detect and trace their orbits due to the much lower contrast between the two objects compared to a stellar primary and brown dwarf companion \citep[e.g.][]{Gelino2011AJ....142...57G, Aberasturi2014AJ....148..129A}. WISE J014656.66+423410.0AB (hereafter W0146) \citep{Dupuy+2015, BardalezGagliuffi2025ApJ...984...74B} is a T9+Y0 binary separated by 0\farcs15 that is currently being monitored to obtain dynamical masses. WISE J033605.05-014350.4AB \citep{Calissendorff_DeFurio2023ApJ...947L..30C} is the first Y-Y dwarf binary detected at a separation of 0\farcs085 with a Y0 primary. The colder companion is likely one of the coldest objects directly detected outside the solar system. Other Y-dwarfs exhibit overluminous behavior that is indicative of multiplicity \citep{Leggett+2021, Albert2025AJ....169..163A}, but their binary status is still unknown, e.g. WISE J053516.80-750024.9, WISEPA J182831.08+265037.8 \citep{DeFurio2023ApJ...948...92D}. CWISEP J193518.59-154620.3 (hereafter W1935) is one such example of an overluminous source relative to the Y-dwarf sequence with suspected multiplicity \citep[Suarez et al. 2025, in prep.,][]{Marocco2019ApJ...881...17M, Kirkpatrick2021ApJS..253....7K}. As a part of GO-2124 (PI: Faherty), W1935 was observed with the James Webb Space Telescope (JWST) NIRSpec/G395H and found to have methane in emission at 3.326 $\mu$m. The authors speculated that auroral processes could explain the feature similar to how it is seen on giant planets.  Given that W1935 was seemingly isolated, the authors also speculated that interactions with an active moon akin to Jupiter interacting with Io, might provide a plausible source for the plasma needed to activate the auroral activity \citep{Faherty2024Natur.628..511F}. In this same program, W1935 was directly imaged with MIRI. In this paper, we present the results of our analysis of the MIRI imaging of W1935 and two known binaries.

In Section \ref{sec:observations}, we describe the data from this program. In Section \ref{sec:dataanalysis}, we describe the method we applied to search for companions to the T and Y dwarfs observed with MIRI. In Section \ref{sec:results}, we present the results of our search and discuss the implications of our findings. In Section \ref{sec:conclusion}, we summarize our results.

\section{Data} \label{sec:observations}
The data used in this paper were observed by the Mid-Infrared Instrument (MIRI) on the \textit{JWST} for GO-2124 (PI: Faherty). W1935 \citep[the Y-dwarf with observed methane emission,][]{Faherty2024Natur.628..511F} was observed on 20 September 2022 with the F1000W, F1280W, and F1800W filters with the FASTR1 readout pattern and two-point dither pattern using 15, 13, and 11 groups per integration for a total exposure time of 83.25, 72.15, and 61.05 seconds, respectively. W1711 \citep[a known T8+T9.5 binary,][]{Liu+2012} was observed on 31 May 2023 with the F1000W, F1280W, and F1800W filters with the FASTR1 readout pattern and two-point dither pattern using 5, 7, and 10 groups per integration for a total exposure time of 27.75, 38.851, and 55.50 seconds, respectively. W0146 \citep[a known T9+Y0 binary,][]{Dupuy+2015} was observed on 21 September 2022 with the F1000W, F1280W, and F1800W filters with the FASTR1 readout pattern and two-point dither pattern using 7, 8, and 10 groups per integration for a total exposure time of 38.85, 44.40, and 55.50 seconds, respectively. All other sources were observed with MIRI between 13 September 2022 and 3 July 2023 with the F1000W, F1280W, and F1800W filters with integration times between 27.75 and 138.75 seconds. We analyzed the stage 2 \textit{cal} files to search for companions. We obtained these data from the Mikulski Archive for Space Telescopes (MAST) at the Space Telescope Science Institute, and can be accessed via \dataset[DOI]{http://dx.doi.org/10.17909/jes9-4v60}.

\section{Data Analysis} \label{sec:dataanalysis}

We performed a point-spread function (PSF) fitting analysis on all the MIRI data using super-sampled empirically derived PSF models \citep{Libralato2023ApJ...950..101L} which we downloaded from \url{https://www.stsci.edu/~jayander/JWST1PASS/LIB/PSFs/STDPSFs/MIRI/}. These models change as a function of position on the detector and the filter in question \citep[e.g.][]{Anderson2000, Anderson2016}. In order to produce a single PSF, we perform the same calculations for each pixel which involves taking the four closest PSF models and performing a bi-cubic interpolation of the portion of those PSF models relevant for each pixel. Then with the resulting pixel values from those four PSFs, we linearly interpolate them based on the proximity of the PSF models to the position on the detector where we are constructing the PSF. 
In order to produce a double-PSF model, we perform the same calculations except this time we produce a second PSF based on the input parameters of the companion, as thoroughly described in \citet{DeFurio2019} with applications to Hubble Space Telescope data.

As performed in \citet{DeFurio2023ApJ...948...92D} and \citet{Calissendorff_DeFurio2023ApJ...947L..30C}, we used the Nested Sampling routine PyMultinest \citep{Buchner2014A&A...564A.125B, Feroz2009MNRAS.398.1601F} to adequately sample the full potential parameter space of companions and identify the best fit single and double PSF solutions to the data in question. In order to take advantage of having multiple images within the same filter and multiple filters of observations, we fit all images simultaneously with the same model accounting for the change in position on the detector due to the two-point dither for both the single and double PSF cases as well as change in filter. Fitting to two images within the F1000W and F1280W filters, our single PSF model has ten parameters: eight parameters for the x and y center of the primary within the first and second images in F1000W and F1280W (i.e. x and y centroid for each image within each filter), and two parameters for flux of the primary in F1000W and F1280W. Our double PSF model has fourteen parameters: eight parameters for the x and y center of the primary within the first and second images in F1000W and F1280W, two parameters for flux of the combined system in both filters, separation of the center of the secondary PSF from that of the primary PSF, position angle of the secondary center relative to the primary center, and two parameters for flux ratio between the secondary and primary PSFs in both filter. We did not include the F1800W images in our analysis which is explained in Sec. \ref{sec:results}. Between the dithered images and different filters, the companion parameters are conserved which may allow for stronger constraints than fitting images independently. We define uniform priors as:  -3 $\leq$ xcen $\leq$ 3 pixels, -3 $\leq$ ycen $\leq$ 3 pixels for each of the four images (two per filter), and -1 $<$ log(flux) $<$ 10 in both filters. Xcen and ycen are adjustments to the input x,y position which we define as the brightest pixel of the source on the detector. For the double PSF model, we keep the same priors for the ten original parameters and set 0.01 $\leq$ separation $\leq$ 4.01 pixels, 0 $<$ position angle $\leq$ 360 degrees, and -5 $\leq$ log(flux ratio) $\leq$ 0 in both filters.


\begin{figure*}[h]
\includegraphics[width=\textwidth]{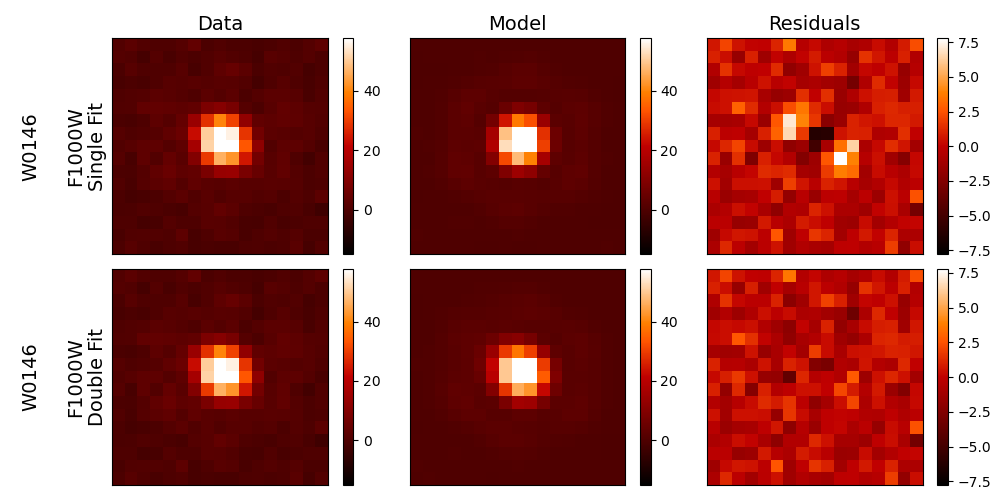}
\includegraphics[width=\textwidth]{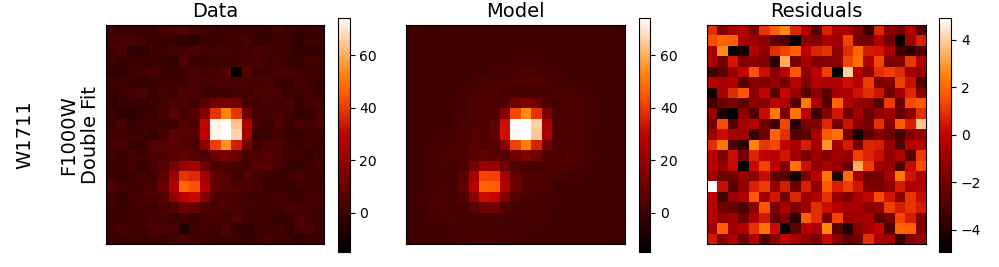}
\caption{As in Figs. \ref{fig:single} and \ref{fig:double}, we simultaneously fit for the companion to W0146 and W1711 across all filters and dither positions. We show an example of one dither position for W0146 in F1000W with the single-PSF fit in the top row and double-PSF fit in the second row. As expected, the residuals follow the pattern for a good model fit to the data, i.e. random noise, for the double fit and structured residuals for the single fit. In the bottom row, we show an example of one dither position for W1711 in F1000W from the double-PSF fit which is also resolvable by eye.}
\label{fig:w0146andw1711}
\end{figure*}

\begin{figure*}[htb]
\includegraphics[width=0.95\textwidth]{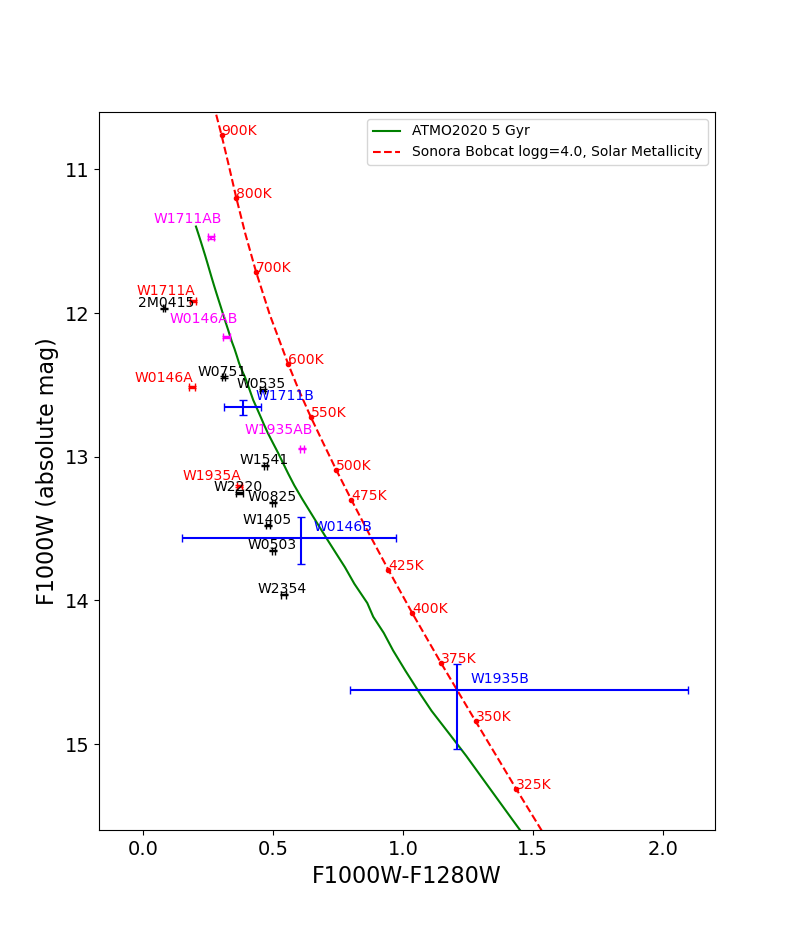}
\caption{F1000W vs. F1000W-F1280W color magnitude diagram for all sources observed with MIRI in GO-2124, overplotted with the ATMO2020 isochrone for 5 Gyr \citep{Phillips+2020} and Sonora Bobcat isochrone for logg=4.0 and solar metallicity \citep{Marley_2021} with select effective temperatures shown in red. All photometry for sources without known companions are shown in black, the combined photometry for W1711, W0146, and W1935 are shown in magenta, the photometry for the primary of each of the three binary systems are shown in red, and the photometry for the companions are in blue. Photometry were taken from Faherty et al. (in prep.). Absolute magnitudes were computed assuming distances from \citet{Kirkpatrick2021ApJS..253....7K}. From only the photometry, W1935B would be the faintest and coldest object observed in this sample with an estimated effective temperature of $\sim$ 350-375 K assuming the Sonora Bobcat models. Nearly all sources seem bluer than the models which may impact their estimated effective temperature and mass.}
\label{fig:isochrone}
\end{figure*}

\section{Results and Discussion} \label{sec:results}
\subsection{WISE1935} \label{subsec:1935}
In Fig. \ref{fig:single} and \ref{fig:double}, we show the results of the best fit single and double PSF models to the F1000W and F1280W data of W1935. The $\chi_{\nu}^{2}$ of the single-PSF fit is 2.3. The residuals demonstrate a clear asymmetric pattern with positive residuals away from the core and negative residuals in the core, indicative of the single-PSF model fitting to the photocenter of a binary system. The $\chi_{\nu}^{2}$ of the double-PSF fit to the four images is 1.4, indicative of data being well-fit by the defined model and demonstrating a clear preference for the double-PSF model over the single-PSF model. We define the significance of the detection based on the $\chi^{2}$ cumulative distribution function with the set number of degrees of freedom and the best fit $\chi_{\nu}^{2}$ values of the single and double PSF fits as shown in eq. 8 of \citet{Gallenne2015}. We calculate a significance of 6.6$\sigma$ showing clear support for the double-PSF fit to the data. When running our code independently on the F1800W images, we do not find a difference between the single-PSF fit ($\chi_{\nu}^{2}$=1.9) and double-PSF fit ($\chi_{\nu}^{2}$=1.8), a 1.16$\sigma$ difference, showing the inability to detect the companion. When included in the simultaneous fitting with the images from the other filters, we can place no constraint on a contrast in the F1800W filter, and we therefore conclude that the companion is unresolved in F1800W.

In Table \ref{tab:companionparameters}, we present the median value of the companion parameters from the posteriors with 1$\sigma$ error bars. The companion is separated by 1.55 pixels or 170.5 milli-arcseconds at a position angle of 150 degrees on the MIRI detector. We use the WCS information from the header of the data to convert from detector space into celestial coordinates and after correcting for the distortion and astrometric solution of MIRI, our best fit separation is 172.2$^{+18.7}_{-7.7}$ milli-arcseconds and position angle is 148.12$^{+2.86}_{-1.98}$ degrees E of N (see Table \ref{tab:companionparameters}). This corresponds to 0.54 $\lambda$/D  and 0.42$\lambda$/D in the F1000W and F1280W filters respectively. This separation in the F1800W filter corresponds to 0.3$\lambda$/D, which is either too close to resolve due to the uncertainty in the PSF models from photon noise or due to the low signal-to-noise of the source in this filter. 
Assuming a distance of 14.43$^{+0.84}_{-0.75}$ pc \citep{Kirkpatrick2021ApJS..253....7K}, this places the companion at a projected separation of 2.48$^{+0.27}_{-0.11}$ au. The companion has a $\Delta$mag = 1.42$^{+0.41}_{-0.18}$ in the F1000W filter and 0.58$^{+0.79}_{-0.37}$ in the F1280W filter. From \citet{Faherty2024Natur.628..511F}, the apparent magnitude of the unresolved system is 13.740 $\pm$ 0.005 and 13.126 $\pm$ 0.007 in the F1000W and F1280W filters, respectively. With the measured contrast, we can estimate the photometry of the primary W1935A as 14.000$^{+0.007}_{-0.007}$ and 13.627$^{+0.007}_{-0.007}$ and of the companion W1935B as 15.420$^{+0.41}_{-0.18}$ and 14.207$^{+0.79}_{-0.37}$ in the F1000W and F1280W filters, respectively.

Using the ATMO2020 models \citep{Phillips+2020} and ages of 1-10 Gyr for field age brown dwarfs, W1935A has masses of 12-39 $M_{\rm Jup}$. For the same ages, W1935B has mass estimates of 7-24 $M_{\rm Jup}$. We expect W1935 to be on the younger end of our age estimate given its proximity (14.43 pc) and Galactic dynamics \citep{Best2024ApJ...967..115B}, meaning the lower mass estimates are more likely to be accurate although not well constrained. This mass range overlaps with masses measured for several directly-imaged low mass exoplanets: e.g. Beta Pictoris b, HR 8799 d and e, and HD 106906b \citep{Marois2008Sci...322.1348M, Lagrange2009A&A...493L..21L,Bailey2014ApJ...780L...4B}. With these estimates, the mass ratio is 0.55-0.62 and W1935B has an effective temperature of $\sim$ 360-420K derived from the ATMO2020 models. Assuming a circular orbit, the period ranges from 16-28 years, making this a valuable system to potentially obtain dynamical masses to calibrate our evolutionary models at the coldest end of the Y-dwarf sequence. From the Sonora Bobcat models \citep{Marley_2021} assuming solar metallicity and logg=4-4.5, W1935B has an estimated effective temperature of 350-375 K. As shown in Fig. \ref{fig:isochrone}, W1935B is the faintest and coldest object in this sample and likely has a spectral type $>$Y1, although follow-up observations are necessary to place strong constraints on its physical parameters.

\subsection{Upper Limit on F1800W Flux of W1935B} \label{subsec:f1800w}
 Our double-PSF fitting code could not detect W1935B in the F1800W filter as stated in Sec. \ref{subsec:1935}. However, we can define an upper limit on the flux at the known position of W1935B by determining the ability of our technique to recover companions at that position as done in \citet{DeFurio2022_BD}. To accomplish this, we created many artificial binaries with the specific position of W1935B, i.e. 1.55 pixels in separations at 150 degree position angle, with an equal flux to the primary, i.e. $\Delta$mag= 0, as the easiest case to recover a companion. The artificial binaries were created from the same empirical PSF models we used in the fitting process for the F1800W filter, normalized to the observed flux of W1935 in F1800W. We created 100 binaries only altering the initial sub-pixel centering across 0.1 pixel bins in x and y space on the MIRI detector. We then added Poisson noise to mimic a true representation of a point source on the detector, and injected this artificial binary onto a clean portion of a MIRI image without any point sources or bad pixels.

We then ran both the single and double PSF-fitting versions of our code on these known binaries. We use the same significance calculation as in Sec. \ref{subsec:1935} to determine if our code was able to detect these injected companions. The maximum difference between the best-fit single and double PSF model is 1.3$\sigma$ with an average difference of 0.8$\sigma$ for all 100 artificial binaries. Therefore, we cannot confidently claim a detection of any of the artificially injected companions although this scenario (equal flux binaries) would be the easiest to recover such a close companion. We cannot place any constraint on the flux of W1935B in F1800W. This could be due to the low signal-to-noise of our observations or due to the broadening of the PSF at longer wavelengths, making the known companion too close to resolve at 18$\mu$m.

\subsection{Other Sources Observed in GO-2124} \label{subsec:other}
While we could not identify any other new companions in the MIRI imaging data, we resolved the previously known companions to W0146 \citep{Dupuy+2015} and W1711 \citep{Liu+2012} in both the F1000W and F1280W filters, but not in the F1800W filter. W1711 appears barely above background in F1800W making the companion undetectable. W0146B is likely too closely separated to resolve in the F1800W filter, separated by 0.26$\lambda$/D, with empirical PSF-fitting or other advanced techniques \citep[c.f. kernel phase interferometry and aperture masking interferometry,][]{Sallum2019JATIS...5a8001S, Kammerer2023PASP..135a4502K, Sallum2024ApJ...963L...2S, Ray2025ApJ...983L..25R}. We present the best fit parameters of the binary systems in Table \ref{tab:companionparameters} and show example fits to these two other brown dwarf binaries in Fig. \ref{fig:w0146andw1711}. Assuming ages of 1-10 Gyr and using the ATMO2020 evolutionary models, we estimate a mass ratio of 0.59-0.67 for W0146 and 0.68-0.72 for W1711 from the F1000W absolute magnitudes. A mass ratio of 0.43-0.54 was estimated for W1711 using J-band photometry in \citet{Liu+2012} which is lower than our estimate. A mass ratio of $\sim$ 0.9 was estimated for W0146 in \citet{Dupuy+2015} which is higher than our estimate. Full spectral coverage of each component in these binary systems is necessary in order to accurately estimate the individual masses. For the tightly separated binaries of W0146 and W1935, orbit monitoring programs to obtain dynamical masses are crucial to calibrate evolutionary models.

In Fig. \ref{fig:isochrone}, we plot the measured photometry of all sources observed with MIRI in GO-2124 (Faherty et al., in prep.). Based on the ATMO2020 and Sonora Bobcat evolutionary models, we see that all sources are bluer than predicted. This points to physical properties in the atmospheres of these extremely cold objects that are not understood in the 10-13 $\mu$m regime and will be discussed further in a companion paper (Faherty et al., in prep). This specific wavelength range is poorly characterized in other brown dwarf programs due to the difficulty in extracting information over these wavelengths using MIRI/LRS \citep{Beiler2024ApJ...973..107B}.

Importantly, both components of all three detected binaries now have photometry consistent with the rest of the late T and Y-dwarf population, compared to the elevated photometry of the unresolved binaries. WISE J053516.80-750024.9 is now the only unresolved single object that lies above the rest of the sequence. This source has been speculated as a binary in the past \citep{Leggett+2021}, and its 10 and 12.8 $\mu$m photometry again may point to the potential for an as of yet unresolved companion.

We see that W0146B has photometry comparable with other Y0-Y1 sources which is consistent with its estimated spectral type of Y0. W1935B is unique in this color-magnitude diagram because it is fainter than of the other late T and Y-dwarfs, indicative of a spectral type $>$Y1.


\section{Implications and Future Work} \label{sec:implications}
Importantly, W1935B is another example of a companion to a late-T or Y-type brown dwarf that does not have a high estimated mass ratio, e.g. W0336 (q=0.56-0.66), J1217 (q=0.40-0.64), W1711 (q=0.43-0.54) \citep{Liu+2012,Calissendorff_DeFurio2023ApJ...947L..30C}. Our past understanding of the mass ratio (q=$M_{2}/M_{1}$) distribution for companions to brown dwarf primaries heavily favored a power law weighted towards q=1 \citep[e.g.][]{Reid2006, Fontanive2018}. With new JWST imaging programs sensitive to extremely cold companions, we have uncovered multiple moderate mass-ratio brown dwarf binaries, suggesting the mass ratio distribution is not as heavily skewed as previously thought. However, this is still based on only a handful of detections. Much time on JWST has been devoted to obtaining spectroscopy of these poorly understood cold objects, but much less time has been devoted to the exploration of multiplicity \citep[e.g.][]{Calissendorff_DeFurio2023ApJ...947L..30C, Albert2025AJ....169..163A, BardalezGagliuffi2025ApJ...984...74B}. Larger programs exploring this new population of cold objects will undoubtedly reveal the presence of more companions and allow us to place strong constraints on the companion population and thus the formation processes that produce these systems.

W1935 is unique in that it recently was shown to have methane emission in its near-infrared spectrum from NIRSpec \citep{Faherty2024Natur.628..511F}. This emission was hypothesized to be caused by an auroral process that may be driven by the presence of an exomoon exciting the upper atmosphere of W1935, i.e. comparable to the effect of Io on Jupiter. However, the detection of a companion now presents a potential alternative hypothesis that the companion (W1935B) is the object with the methane emission caused by a still unknown auroral process. Other mechanisms due to the presence of a companion may drive plasma formation, see \citet{Kao2025MNRAS.539.2292K}. In order to attempt to disentangle the methane emission feature, observations with NIRSpec/IFU spectroscopy may allow us to extract the spectrum of both components within this system and determine which component has the methane emission and which hypothesis is most likely. Spectroscopic observations over a long time baseline would allow us to identify any periodicity in the methane emission. NIRSpec fixed slit prism observations are planned for W1935 in Cycle 4 (GO-7793, PI: Faherty). Alternatively, we can also probe the exomoon hypothesis by tracing the orbit of the companion and determining if a moon around either component could have a stable configuration. This system is unique both in that it has methane in emission and is only the second Y-Y dwarf binary ever detected. This new object also helps fill in the gap in the Y-dwarf sequence between the Y0-Y1 objects and the extremely cold W0855, the only Y4 ever detected \citep{Kirkpatrick2019, Luhman2024AJ....167....5L}.

\begin{deluxetable*}{cccccccc}
\tablenum{1}
\tablecaption{Best-fit companion parameters for W1935B, W0146B, and W1711B and photometry for each component in the binary system in apparent magnitudes with 1$\sigma$ errors. We also include a mass ratio estimate for each binary system using the ATMO2020 evolutionary models, the absolute magnitude in F1000W, and assuming ages 1-10 Gyr. \label{tab:companionparameters} }

\tablewidth{0pt}
\tablehead{
\colhead{Source}	& \colhead{Separation}	&	\colhead{Position Angle}	&	\colhead{$\Delta$mag}	&	\colhead{$\Delta$mag} & \colhead{F1000W} & \colhead{F1280W} & \colhead{Mass Ratio}\\
\colhead{}	& \colhead{(mas)}	&	\colhead{(degrees E of N)}	&	\colhead{(F1000W)}	&	\colhead{(F1280W)} & \colhead{(Vega mag)} & \colhead{(Vega mag)}  & \colhead{Estimate}}

\startdata
W1935A &	&	& &	& 14.000$^{+0.007}_{-0.007}$ & 13.631$^{+0.007}_{-0.007}$    \\
W1935B & 172.2 $^{+18.7}_{-7.7}$	&	148.12$^{+2.86}_{-1.98}$	&	1.42$^{+0.41}_{-0.18}$ &	0.58$^{+0.79}_{-0.37}$	 & 15.420$^{+0.41}_{-0.18}$ & 14.211$^{+0.79}_{-0.37}$ & 0.55-0.62 \\
\hline
W0146A &	&	& &	& 13.949$^{+0.007}_{-0.007}$ & 13.763$^{+0.01}_{-0.01}$    \\
W0146B & 149.9 $^{+26.4}_{-7.7}$	&	273.25$^{+3.86}_{-2.23}$	&	1.05$^{+0.18}_{-0.15}$	& 0.63$^{+0.32}_{-0.43}$ & 14.999$^{+0.18}_{-0.15}$ & 14.393$^{+0.32}_{-0.43}$	& 0.59-0.67 \\
\hline
W1711A &	&	& &	& 13.734$^{+0.008}_{-0.008}$ & 13.542$^{+0.009}_{-0.009}$   \\
W1711B & 701.6 $^{+9.9}_{-6.6}$	&	341.65$^{+2.23}_{-0.62}$	&	0.74$^{+0.05}_{-0.05}$	   & 0.55$^{+0.05}_{-0.05}$ & 14.474$^{+0.05}_{-0.05}$ & 14.092$^{+0.05}_{-0.05}$  &  0.68-0.72  \\
\enddata
\end{deluxetable*}

\section{Conclusion} \label{sec:conclusion}

We analyzed MIRI imaging data for a set of late-T and Y-dwarfs and found a new, colder companion to the Y-dwarf W1935A. The companion was detected in the F1000W and F1280W filters at a separation of 1.55 pixels (172.2 mas after correction for the distortion and astrometric solution, 0.54$\lambda$/D and 0.42$\lambda$/D) or 2.48 au assuming a distance of 14.43 pc. We estimate a mass of 7-24 $M_{\rm Jup}$ for W1935B and 12-39 $M_{\rm Jup}$ for W1935A assuming ages from 1-10 Gyr for a mass ratio of 0.55-0.62. This new object is likely one of the coldest compact objects ever detected outside the solar system, c.f. 14 her c \citep{BardalezGagliuffi2025arXiv250609201B}, bridging the gap in the Y-dwarf sequence, although follow-up observations are necessary to characterize the spectral energy distribution and derive the dynamical mass of this newly discovered Y-dwarf companion. 

\section*{Acknowledgments}
This work is based on observations made with the NASA/ESA/CSA James Webb Space Telescope. The data were obtained from the Mikulski Archive for Space Telescopes at the Space Telescope Science Institute, which is operated by the Association of Universities for Research in Astronomy, Inc., under NASA contract NAS 5-03127 for JWST. These observations are associated with program No. 2124. J.M.V. acknowledges support from a Royal Society - Research Ireland University Research Fellowship (URF/1/221932, RF/ERE/221108). M.D.F. is supported by an NSF Astronomy and Astrophysics Postdoctoral Fellowship under award AST-2303911. M.D.F. acknowledges S.F.

\bibliography{references_thesis}{}
\bibliographystyle{aasjournal}

\end{document}